# X-ray mirror figure correction using differential deposition

Ch. Morawe, S. Labouré, F. Perrin, A. Vivo, R. Barrett

ESRF, Grenoble, France

## ABSTRACT

The surface figure of x-ray mirrors can be improved by differential deposition of thin films. To achieve the required corrections, $WSi_2$ layers of variable thickness were deposited through beam-defining apertures of different openings. The substrates were moved in front of the particle source with specific velocity profiles that were calculated with a deconvolution algorithm. Two different DC magnetron sputter systems were used to investigate the correction process. Height errors were evaluated before and after each iteration using off-line visible light surface metrology. Four 300 mm long flat Si mirrors were used to study the impact of the initial shape errors on the performance of the correction approach. The shape errors were routinely reduced by a factor of 20-30 down to levels below 0.5 nm RMS.

## 1. INTRODUCTION

Surface figure errors critically limit the performance of reflective X-ray optics since they alter the shape of the reflected wave front. In modern hard X-ray light sources with increasing spatial coherence [1], figure accuracies below 1 nm are required to preserve the source properties [2].

To correct for figure errors precise metrology and deterministic polishing techniques such as elastic emission machining (EEM) [3], ion beam figuring (IBF) [4-6], and differential deposition (DD) have been developed. EEM and IBF rely on material removal, DD adds material by thin film coating. It has been developed within the astronomy community [7], but has also been applied to synchrotron optics [8-11].

Following previous work [12-15] carried out on a Large Multilayer Coating System (LMCS) [16], the DD capabilities of the recently installed Compact Multilayer Coating System (CMCS) [17] were studied. A series of 300 mm long flat Si mirrors was treated, whose surface figures were corrected first using the LMCS and then employing the CMCS.

## 2. METHOD

### 2.1 Theory

The differential deposition technique is based on a substrate that moves in front of a particle source following a specific velocity profile $v(x_m)$. The resulting film thickness distribution $t(x_s)$ can be expressed as [18]

$$t(x_s) = \int_{-S}^{+S} R \cdot f(x_m - x_s) \cdot \frac{dx_m}{v(x_m)} \quad (1)$$

where $f(x_m-x_s)$ is the normalized static particle flux profile of the source on the substrate and $R$ the growth rate at its centre. $x_s$ and $x_m$ are the substrate and motion coordinates, respectively. The integration is carried out over the full length $2S$ of the substrate motion. In the present case, the thickness distribution $t$ is given and the speed profile $v$ needs to be calculated, which corresponds to a deconvolution process. Using discrete steps equation (1) can be written as a system of linear equations. An algorithm based on matrix inversion was developed to solve the problem. It includes sub-routines that were originally developed for astronomical image deconvolution by NASA [19]. The growth rate $R$ and the flux profile $f$ are experimental input parameters for the programme.

### 2.2 Thin film deposition

The coatings were made at the ESRF multilayer laboratory [16,17] using DC magnetron sputtering. The deposition process took place in an Ar atmosphere at a working pressure of 0.1 Pa. $WSi_2$ was deposited from compound targets. Beam defining apertures with openings of 24 mm / 2 mm on the LMCS and 10 mm / 1 mm on the CMCS were placed in front of the substrate surface. Stationary WSi$_2$ coatings, where the thickness is controlled by opening and closing a shutter, were carried out to measure the particle flux distribution in the substrate plane. The corrective coatings were made in dynamic mode where the substrate moves in front of the aperture following a pre-programmed speed profile. Repeated duty cycles were applied to obtain the required thickness profile. The principal process parameters of both deposition machines are summarized in Table 1.

| Deposition machine | LMCS | CMCS |
|---|---|---|
| Power [W] | 200 | 50 |
| WSi$_2$ growth rate [nm/s] | 0.31-0.33 | 0.40-0.50 |
| Distance target-substrate [mm] | 92 | 81 |
| Distance aperture-substrate [mm] | 3 | 1 |
| Sample position accuracy [mm] | ±0.3 | ±0.1 |

Tab.1: Principal process parameters of the deposition machines.

### 2.3 X-ray reflectivity

Test coatings were characterized on a laboratory X-ray reflectometer at 8048eV [20]. Specular reflectivity scans can be carried out with a dynamical range of up to $10^7$. Simulation software based on the Parratt formalism [21] allows for the precise determination of thicknesses, mass densities, and interface widths.

### 2.4 Surface metrology

Both static thickness profiles and surface figures were measured with a ZYGO phase-shifting Fizeau interferometer at 633 nm with 150 mm diameter aperture. In order to extend the measurement capability to mirrors up to 1 meter long, stitching techniques have been developed and implemented [22,23] on this instrument. Acquisition of overlapped sub-apertures is ensured by a motorized stitching tool composed of translation, tip-tilt and rotation stages allowing a fine alignment of the horizontally reflecting mirror while the interferometer stays at a fixed position. The set of sub-apertures is then stitched using PyLost (Python Large Optic Stitching) software [24] to reconstruct the topography of the entire mirror surface. The repeatability achieved is better than 0.05 nm RMS and cross comparison with Long Trace Profiler data has confirmed a sub-nanometer accuracy. For LTP measurements, mirrors under test are supported by 3 spheres spaced by half of its length in order to subtract gravity effects and allow comparison with Fizeau data. Gravity is not negligible for these 300 mm long and only 15 mm thick mirrors, inducing height errors of 4.2 nm RMS (13.3 nm PV) in case of <100> Silicon orientation. The residual shape error profile is obtained by subtracting a second order polynomial from the measured height profile. In this work, overall shape errors are referred to as Root Mean Square (RMS) and Peak to Valley (PV).

The surface roughness was investigated using a Wyko NT9300 white light interferometer. Three parfocal objectives (50X, 5X, and 2.5X) were used on an automated turret offering respective optical resolutions from 0.5 µm to 3.8 µm and fields of view (FOV) between 0.126 x 0.095 mm$^2$ and 2.50 x 1.90 mm$^2$. The instrument is also equipped with encoded and motorized X and Y translation stages. The accuracy of the translation allows tracking of the same measurement point after repositioning the mirror, provided two fiducial markers can be identified, thus allowing to follow up rigorously the process of differential deposition with respect to micro-roughness in a spatial frequency domain from 0.4 mm$^{-1}$ to 2 µm$^{-1}$.

## 3. RESULTS

### 3.1 Static flux distribution

To characterize the flux distribution thin WSi$_2$ films were deposited on stationary substrates through apertures of 24 mm / 2 mm (LMCS) and 10 mm / 1 mm (CMCS). Their local thickness was measured with XRR and Fizeau interferometry. The normalized thickness profiles are shown in Fig.1. The resulting widths vary between 25.0 mm and 1.4 mm FWHM. This choice offers the possibility to initially correct long period figure errors faster using a broad flux profile and to refine the process with further iterations selecting narrower apertures.

The LMCS profiles show two pronounced shoulders that are edge projections of the two straight erosion lines on the rectangular sputter target. This effect is attenuated on the CMCS where circular sputter targets are used. The CMCS profiles show steeper edges than on the LMCS since the respective apertures are approached closer to the substrate surface.

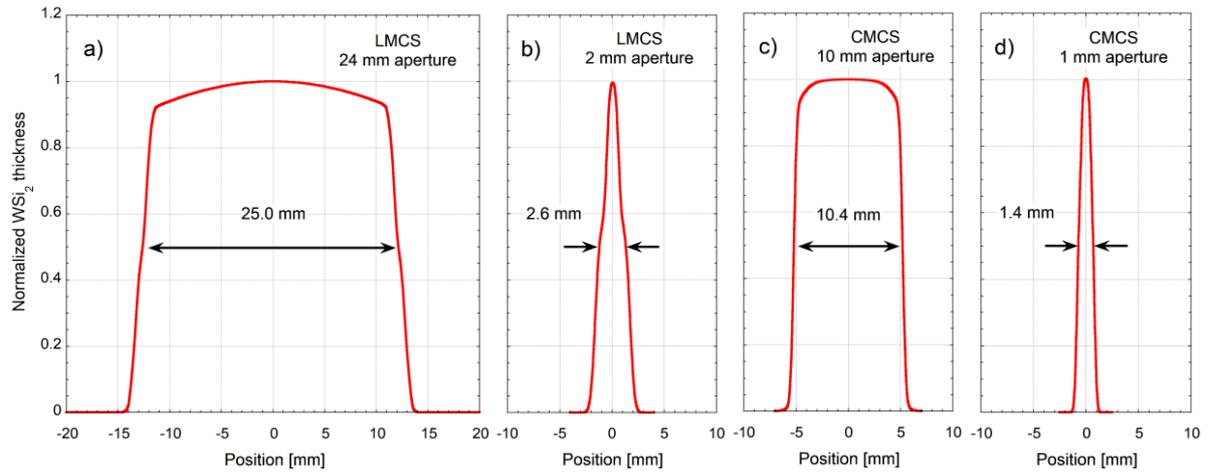

Fig.1: Normalized thickness profiles of $WSi_2$ films coated on stationary substrates through apertures of 24 mm (a) and 2 mm (b) on the LMCS and of 10 mm (c) and 1 mm (d) on the CMCS.

### 3.2 Figure correction of Si mirrors

A series of four 300 mm long, 40 mm wide, and 15 mm thick flat Si mirrors was selected for correction. Both the surface metrology and the differential deposition technique were applied to the central line along a trace of 280 mm. For practical reasons, all used metrology data sets were resampled to 0.5 mm steps and the corresponding speed profiles were applied with the same step size. This approach appears justified since the experimental flux profiles do not allow for corrections of features smaller than about 2 mm. To provide identical surface conditions during all subsequent metrology studies, the Si substrate was first covered with a 50 nm thick uniform $WSi_2$ film deposited on the LMCS without beam defining aperture. Along with each corrective coating a 50 nm thick uniform $WSi_2$ layer was added to attenuate potential parasitic signals from buried oxide layers that may distort the metrology results [14]. These typically form when the mirrors are removed from the vacuum. The references of the mirrors under study are WP#30, WP#32, WP#37, and CO#7. The applied correction protocol is summarized in Table 2.

| Iteration | Machine | Aperture [mm] | Coating [nm] |
|---|---|---|---|
| #0 | - | - | Si substrate |
| #1 | LMCS | 100 | $WSi_2(50)$ |
| #2 | LMCS | 24 | $WSi_2(50)/WSi_2(var)$ |
| #3 | LMCS | 2 | $WSi_2(50)/WSi_2(var)$ |
| #4 | CMCS | 1 | $WSi_2(50)/WSi_2(var)$ |

Tab.2: Correction protocol applied to all four Si mirrors.

As an example, Fig.2 shows the evolution of the 2-D surface error profile of mirror WP#30, measured with the Fizeau interferometer. The images are horizontally compressed. Each FOV covers a length of 280 mm and a width of 10 mm (7 mm for images #3 and #4). The vertical numbering from #1 to #4 corresponds to the iteration sequence defined in Tab.2. Note that the image of iteration #3 appears twice as the height colour scale is refined for better visibility of the final correction steps. In all cases, the DD correction was applied to the central horizontal line profile of each image.

The first corrective coating (iteration #1 to #2) using the 24 mm aperture removed the dominating long period variations while the second correction (iteration #2 to #3) through the 2 mm aperture attenuated medium period features, in particular those near the centre. The third correction (iteration #3 to #4) with the 1 mm aperture reduced the short period errors down to the intrinsic limit given by the flux profile. Apart from the general improvement one observes that, as expected, the shape optimization is focused on the central horizontal trace along the mirror. Below and above this line, the figure error increases. One can also notice localized surface defects that cannot be corrected for with the given technique.

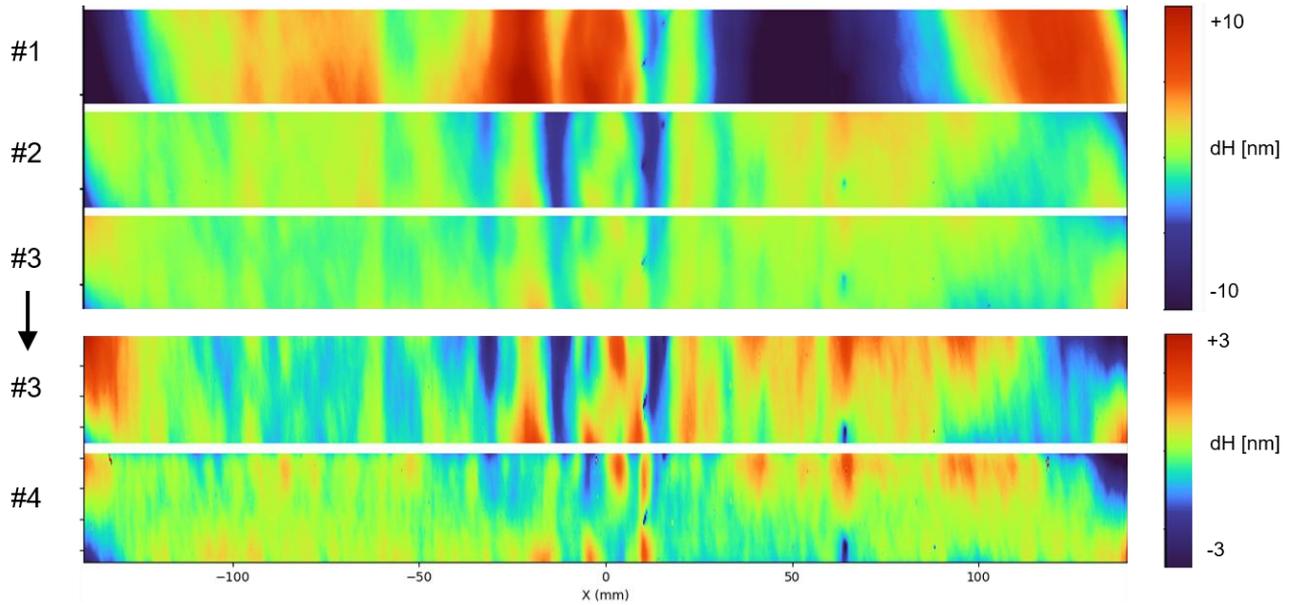

Fig.2: 2-D surface profile of mirror WP#30 measured with the Fizeau interferometer. Each FOV covers a length of 280 mm and a width of 10 mm (7 mm for #3 and #4). The numbering from #1 to #4 corresponds to the iteration sequence defined in Tab.2. The colour height scales are added on the right of the respective images.

Figure 3 summarizes the shape errors of mirror WP#30 averaged over a 1 mm wide central horizontal line taken from the Fizeau 2-D data. The stepwise improvements are clearly visible. Over the analysis length of 280 mm, the surface figure error was reduced from 5.52 nm RMS and 24.29 nm PV (black curve) to 0.20 nm RMS and 1.52 nm PV (green curve).

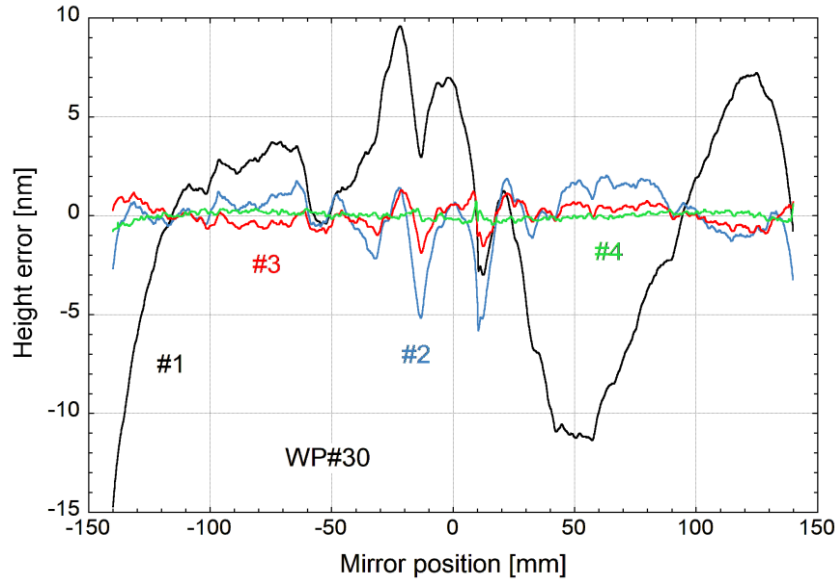

Fig.3: Fizeau measured surface height errors of mirror WP#30 for iterations #1 to #4 as given in Tab. 2.

Figure 4 shows the residual shape errors of all four mirrors after iteration #4 and on a common scale. The curves contain essentially short period modulations and spikes that could not be corrected for with the available hardware. The underlying long period variations are difficult to interpret since they may be affected by the choice of the second order polynomial that is subtracted from the raw data. These long period undulations have a weaker impact on the deformation of the wave front upon reflection than short period oscillations. Note that the principal residual errors stay below 1 nm on all 4 mirrors.

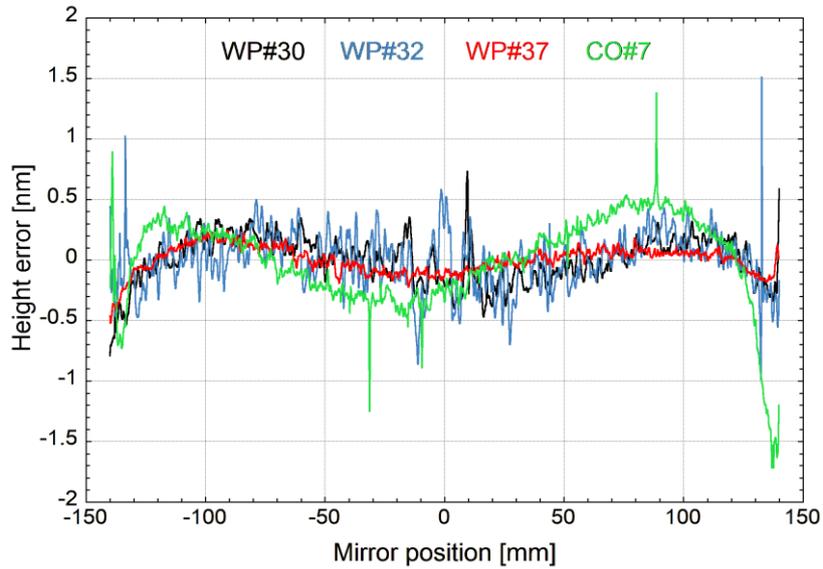

Fig.4: Fizeau measured residual surface height errors of all four mirrors after iteration #4.

The surface figure errors were analysed in the same way for all four mirrors. The evolution versus iteration steps over the full length of L = 280 mm is summarized in Fig.5a. The clear aperture of these 300 mm long mirrors is typically only 260 mm, since they are being used on dynamic bending systems. Consequently, the corresponding data sets were re-examined over a reduced length of 260 mm, as shown in Fig.5b. Iteration #0 was omitted since there were no significant figure error changes after the initial uniform coating. RMS errors are indicated by dots, PV errors by squares. All height errors are plotted on a logarithmic scale.

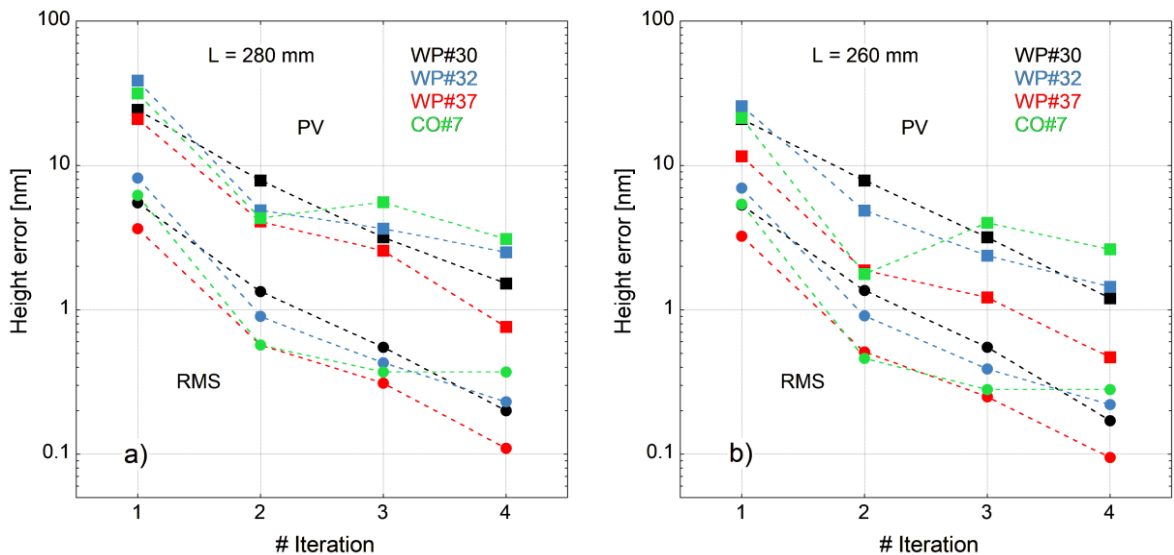

Fig.5: Surface figure errors of mirrors WP#30 (black), WP#32 (blue), WP#37 (red), and CO#7 (green) versus iteration steps over analysis lengths of 280 mm (a) and 260 mm (b). RMS errors are indicated by dots, PV errors by squares. Broken lines are guides to the eye.

As expected, the RMS error levels are nearly one order of magnitude below those of the PV values. They are less affected by the shortening of the analysis length that primarily cuts down the PV values near the mirror edges. For all three WP mirrors one observes a continuous and nearly exponential decay with some spread of both the RMS and the PV errors. This means that each iteration contributes to a significant improvement of the mirror surface figure. Mirror WP#37 reaches the best residual error levels with 0.1 nm RMS and 0.5 nm PV. For mirror CO#7, the RMS error appears to level off after iteration #3 and the PV error slightly increases. This can be explained by a pronounced level of pollution that appeared on this mirror after step #3. The corresponding numerical values are listed in Table 3. The total improvement factors were added in the last row.

| Mirror | WP#30 | | WP#32 | | WP#37 | | CO#7 | |
|---|---|---|---|---|---|---|---|---|
| dH [nm] | RMS | PV | RMS | PV | RMS | PV | RMS | PV |
| #1 | 5.52 (5.31) | 24.29 (20.94) | 8.18 (6.96) | 38.79 (25.70) | 3.65 (3.24) | 20.93 (11.59) | 6.22 (5.40) | 31.62 (21.38) |
| #2 | 1.34 (1.36) | 7.87 (7.87) | 0.90 (0.91) | 4.88 (4.86) | 0.57 (0.51) | 4.10 (1.87) | 0.57 (0.46) | 4.32 (1.77) |
| #3 | 0.55 (0.55) | 3.18 (3.18) | 0.43 (0.39) | 3.64 (2.37) | 0.31 (0.25) | 2.57 (1.22) | 0.37 (0.28) | 5.56 (4.00) |
| #4 | 0.20 (0.17) | 1.52 (1.20) | 0.23 (0.22) | 2.50 (1.44) | 0.11 (0.095) | 0.76 (0.47) | 0.37 (0.28) | 3.10 (2.63) |
| Improvement factor | 27.6x (31.2x) | 16.0x (17.5x) | 35.6x (31.6x) | 15.5x (17.8x) | 33.2x (34.1x) | 27.5x (24.7x) | 16.8x (19.3x) | 10.2 (8.1x) |

Tab.3: List of height errors as shown in Fig.5a measured over a length of 280 mm. Values plotted in Fig.5b covering a length of 260 mm are added in brackets. Total improvement factors are given in the last row.

### 3.3 Power spectral density

An alternative way to analyze the height error evolution is to transform the data into frequency space and to derive corresponding power spectral densities (PSD). 1-D PSDs were calculated using standard routines available in the SciPy module [25] and applied to 1 mm averaged profiles extracted from the 2-D Fizeau data. A Tukey (tapered cosine) window with a ratio of taper to constant sections of 0.2 and a total width of 280 mm is applied prior to PSD calculation in order to reduce frequency leakage phenomena due to the extremities of the measured area [26]. The cumulative power spectral density (CPSD) is the integral over frequency of the one-sided line PSD and provides a graphic representation of the surface variance build-up with increasing frequency. The square root of this function represents the RMS height error build-up and shows rather clearly the contribution of different spatial frequencies to the overall shape errors. Figure 6 present the PSD (a) and the corresponding $(CPSD)^{1/2}$ (b) of the height error profiles measured on mirror WP#30, after iterations #1 to #4. The frequency cut-off near 5 mm$^{-1}$ is given by the Fizeau sampling interval of about 0.1 mm.

In Fig.6, the black curves of iteration #1 represent the situation after the initial uniform $WSi_2$ coating. After iteration #2 (blue curves) there is a clear signal reduction in the spatial frequency range below 0.03 mm$^{-1}$, corresponding to the spatial wavelength range above 30 mm, which agrees with the 24 mm wide aperture used for this correction. The spectra after iteration #3 (red curves) indicate an additional reduction up to frequencies of 0.2 mm$^{-1}$ or down to periods of 5 mm, which is linked to the use of the 2 mm aperture. After iteration #4 (green curves) this limit is pushed further to 0.5 mm$^{-1}$ or 2 mm due to the insertion of the 1 mm slit. All corrections efficiently reduce the surface figure errors in spatial frequency ranges that correspond to the respective aperture widths.

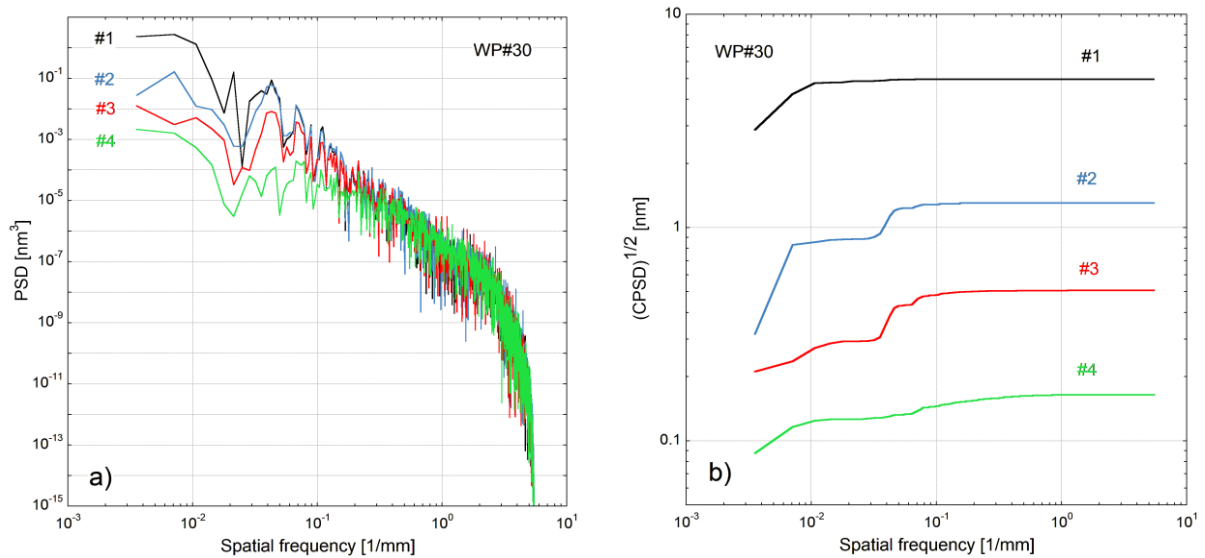

Fig.6: PSD (a) and $(CPSD)^{1/2}$ spectra (b) of mirror WP#30 calculated from Fizeau data for iterations #1 to #4, both given on double logarithmic scales.

### 3.4 Roughness evolution

Previous studies [13] have shown that the surface roughness of $WSi_2$ films on Si is lower than what was found for metallic layers such as Cr or Pt and that it remains virtually constant up to a thickness of about 300 nm. In the present context, it was investigated how the roughness evolves after each iteration step that was required to correct the figure errors. Figure 7 shows the 15-point average RMS surface roughness versus iteration steps measured with the 5X objective (a) and with the 50X objective (b) of the Wyko interferometer. For mirror CO#7, one zone was omitted, because it was covered by a particle. Mirrors WP#30, WP#37, and CO#7 maintain very low roughness levels of about 0.1 nm RMS on both length scales. On mirror WP#32 (blue dots) there is a progressive roughness increase in the course of the DD treatment. This mirror had been clamped in a bender in the past and the affected zones near the edges piled up roughness as coatings were added.

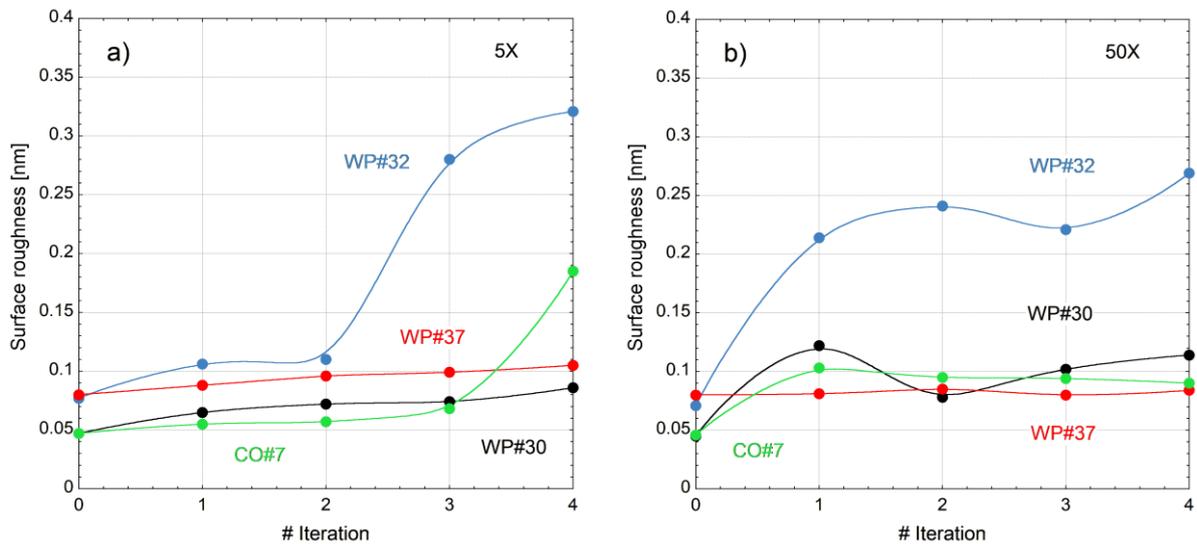

Fig.7: RMS surface roughness versus iteration steps of mirrors WP#30 (black dots), WP#32 (blue dots), WP#37 (red dots), and CO#7 (green dots). Fig.7a shows data obtained with objective 5X, Fig.7b with 50X. Solid lines are guides to the eye.

## 4. DISCUSSION AND CONCLUSIONS

The figure error correction with DD of 4 Si mirrors provided an opportunity to study the accuracy and the repeatability of the process both on the deposition and on the metrology side. The mirrors had similar initial average figure errors but different individual shapes. The achieved figure improvement for the WP mirrors is about 30x (RMS) and 20x (PV) with relatively little spread. Mirror CO#7 scores less well due to surface pollution. This work shows that residual figure error levels of 0.1 nm RMS and 0.5 nm PV can be reached over a 260 mm long clear aperture and down to spatial periods of 2 mm. This result is important in view of applying DD routinely for the correction of synchrotron beamline mirrors and multilayers.

It was confirmed that parasitic reflections can be avoided by adding a 50 nm thick uniform $WSi_2$ layer at each iteration. The newly employed CMCS clearly facilitates the surface figure error reduction towards higher spatial frequencies both due to narrower apertures and thanks to a more accurate sample positioning and motion system.

Figure 8 shows the time evolution of the RMS height errors achieved on different 300 mm long Si mirrors with DD and using different layer materials, all carried out at the ESRF, since the process was first tested in 2018. There is a significant improvement of nearly one order of magnitude over 6 years or a gain of 2x every 1.8 years, according to an exponential fit (red line). It is instructive to compare this time series with Fig.4 in reference [6], where similar considerations were made in the context of IBF.

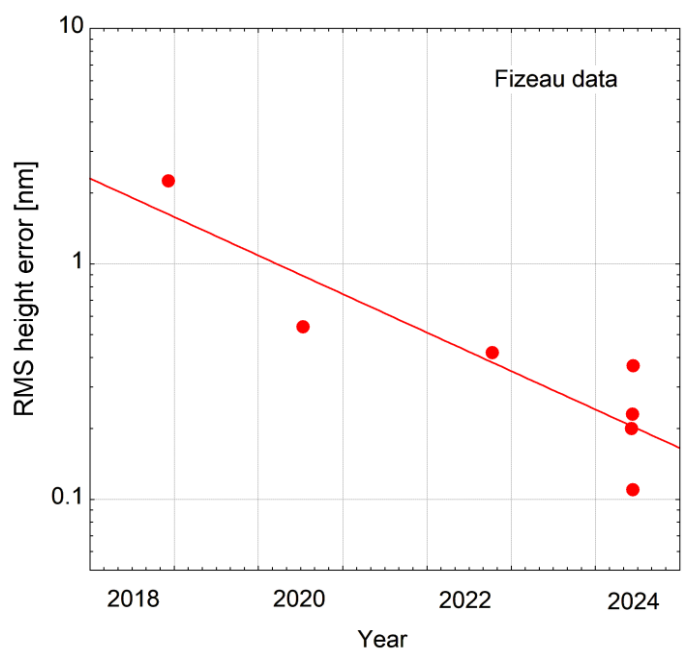

Fig.8: Time evolution of RMS height errors achieved on different 300 mm long Si mirrors with DD at the ESRF. The solid line is an exponential fit to the data.

Due to the recent improvements of both the deposition process and the metrology protocol, the exact and reproducible sample positioning in both environments has become increasingly critical. More efforts will be necessary to implement advanced techniques such as the use of fiducials and references to minimize potential offsets.

On the CMCS there is room for further reduction of the aperture size or for the implementation of beam collimation to generate even narrower flux profiles. Technical solutions for in-situ metrology are under study. This would allow for faster convergence during the initial correction steps where the measurement accuracy is less demanding. It would also reduce the risk of contamination during sample transfer and exposure.

Subsequent multilayer coatings on DD corrected mirrors need to be carried out and tested under operational conditions on 4$^{th}$ generation synchrotron beamlines to validate the maturity of this technology.

## ACKNOWLEDGEMENTS

This project has received funding from the European Union's Horizon 2020 research and innovation programme under grant agreement No. 101004728.